\documentclass[12pt]{article}
\usepackage{epsf}
\usepackage{amsmath}
\usepackage{graphicx}
\usepackage{color}
\usepackage{psfrag}
\usepackage{amssymb}
\usepackage{bm,epsfig,amsmath,amssymb,amsfonts,
latexsym,comment,cancel,verbatim}
\usepackage[font=md,captionskip=8pt]{subfig}
\usepackage{amsfonts}
\usepackage{epsfig}

\usepackage{ifpdf}

\renewcommand{\baselinestretch}{1.3}
\newcommand{\bmat}{\left(\begin{array}}
\newcommand{\emat}{\end{array}\right)}

\def\yzero{\smash{\hbox{$y\kern-4pt\raise1pt\hbox{${}^\circ$}$}}}

\def\beq{\begin{equation}}
\def\eeq{\end{equation}}
\def\beqa{\begin{eqnarray}}
\def\eeqa{\end{eqnarray}}

\def\-{\hphantom{-}}

\def\s2{\frac{1}{\sqrt2}}

\def\beq{\begin{equation}}
\def\eeq{\end{equation}}
\def\beqa{\begin{eqnarray}}
\def\eeqa{\end{eqnarray}}

\def\Tr{{\rm Tr \,}}

\def\nn{\nonumber}

\def\IF{\relax{\rm I\kern-.18em F}}
\def\II{\relax{\rm I\kern-.18em I}}

\def\Dsl{\,\raise.15ex\hbox{/}\mkern-13.5mu D} 

\catcode`\@=11   
\newdimen\@rotdimen
\newbox\@rotbox  

\def\@vspec#1{\special{ps:#1}}
\def\@rotstart#1{\@vspec{gsave currentpoint currentpoint translate
   #1 neg exch neg exch translate}}
\def\@rotfinish{\@vspec{currentpoint grestore moveto}}
\def\@rotr#1{\@rotdimen=\ht#1\advance\@rotdimen by\dp#1%
   \hbox to\@rotdimen{\hskip\ht#1\vbox to\wd#1{\@rotstart{90 rotate}%
   \box#1\vss}\hss}\@rotfinish}
\def\@rotl#1{\@rotdimen=\ht#1\advance\@rotdimen by\dp#1%
   \hbox to\@rotdimen{\vbox to\wd#1{\vskip\wd#1\@rotstart{270 rotate}%
   \box#1\vss}\hss}\@rotfinish}%
\def\@rotu#1{\@rotdimen=\ht#1\advance\@rotdimen by\dp#1%
   \hbox to\wd#1{\hskip\wd#1\vbox to\@rotdimen{\vskip\@rotdimen
   \@rotstart{-1 dup scale}\box#1\vss}\hss}\@rotfinish}%
\def\@rotf#1{\hbox to\wd#1{\hskip\wd#1\@rotstart{-1 1 scale}%
   \box#1\hss}\@rotfinish}%
\def\rotate{\@ifnextchar[{\@rotate}{\@rotate[l]}}
\def\@rotate[#1]#2{\setbox\@rotbox=\hbox{#2}\@nameuse{@rot#1}\@rotbox}

\catcode`\@=12

\topmargin
-2cm
\textwidth
16.5cm
\textheight
24cm
\oddsidemargin
0cm
\evensidemargin
0cm

\begin{document}

\makeatletter
\@addtoreset{equation}{section}
\makeatother
\renewcommand{\theequation}{\thesection.\arabic{equation}}

\pagestyle{empty}
\rightline{IFT-UAM/CSIC-11-81} \rightline{FTUAM-11-61}
\vspace{0.5cm}
\begin{center}
\LARGE{\bf   The MSSM Higgs Sector with a\\ Dynamical Goldstino Supermultiplet  \\[20mm]}
 \large{C}\normalsize{HRISTOFFER} \large{P}\normalsize{ETERSSON}${}^{\mathrm{a}}$, \large{A}\normalsize{LBERTO} \large{R}\normalsize{OMAGNONI}${}^{\mathrm{a,b}}$\\[10mm]
\footnotesize{${}^{\mathrm{a}}$\emph{Instituto de F\'{\i}sica Te\'orica (UAM/CSIC)}}\\
\footnotesize{${}^{\mathrm{b}}$\emph{Departamento de F\'{\i}sica Te\'orica (UAM)}}\\[1.7mm]
\footnotesize{\emph{Universidad Aut\'onoma de Madrid, Cantoblanco, 28049 Madrid, Spain}}\\[10mm] 
\emph{Email}: $\textsf{ christoffer.petersson@csic.es,  alberto.romagnoni@uam.es}$ \\[10mm] 
\small{\bf Abstract} 
\end{center}

\noindent We consider a supersymmetric realization of the MSSM Higgs sector, where the soft terms are promoted to supersymmetric operators and a minimal weakly coupled hidden sector is included. The model exhibits long-lived meta-stable vacua in which supersymmetry and electroweak symmetry are spontaneously broken. The spectrum contains, in addition to the usual MSSM particles and the goldstino fermion, a CP-even and a CP-odd scalar in the neutral Higgs sector corresponding to the complex sgoldstino scalar. By treating all the components of the goldstino supermultiplet dynamically and taking into account their interactions with the Higgs fields, additional couplings beyond those of the MSSM are induced. When the supersymmetry breaking scale is low, these couplings can raise the masses of all the Higgs particles above the LEP bound, already at tree level and for any value of $\tan\beta$. The model includes a scenario where, for any choice of the supersymmetry breaking scale, the set of parameters is reduced to the standard set $(\mu, B_{\mu}, \tan \beta)$ of the MSSM Higgs sector but where novel decays of doublet-like states into sgoldstino-like states are kinematically allowed.

\newpage
\setcounter{page}{1}
\pagestyle{plain}
\renewcommand{\thefootnote}{\arabic{footnote}}
\setcounter{footnote}{0}


\section{Introduction}

Supersymmetric extensions of the standard model (SM) have the potential to stabilize the weak scale, dynamically explain why the weak scale is hierarchically smaller than the Planck scale, allow for gauge coupling unification and provide a viable dark matter candidate. However, there is currently no conclusive argument that selects a particular scale or mechanism for supersymmetry breaking and it is in general difficult to avoid fine-tuning. In the context of the minimal supersymmeric SM (MSSM) there is the so called ``little hierarchy problem''  stemming from the fact that the MSSM tree level prediction for the mass of the lightest Higgs particle is bounded from above by the mass of the $Z$-boson. Radiative corrections 
can raise it above the LEP bound of 114 GeV \cite{Barate:2003sz}, but at the expense of separating the weak scale from the mass scale of the superpartners (in particular the top squark masses\footnote{It is also possible to increase the Higgs mass by considering highly mixed top squarks.}), to which the mass of the lightest Higgs particle is logarithmically sensitive. The problem is that, since the Higgs potential depends quadratically on the superpartner mass scale, such a separation requires unnatural cancellations between Higgs parameters in order for the potential to give rise to an electroweak symmetry breaking vacuum expectation value (VEV) of 174 GeV.

The little hierarchy problem can be interpreted as a hint for non-minimal supersymmetric extensions of the SM containing degrees of freedom beyond the MSSM ones. A model-independent way of parametrizing the effect of such additional degrees of freedom and to raise the tree level mass of the lightest  Higgs particle is to add effective operators involving the Higgs fields \cite{Brignole:2003cm} (see also \cite{extraops} for more recent discussions). In terms of specific models, a well-studied one is the next to MSSM (NMSSM) (see \cite{Ellwanger:2009dp} for an extensive review) where the degrees of freedom of a gauge singlet chiral superfield is added to the MSSM spectrum.  Even though the NMSSM is designed to solve the ``$\mu$-problem'' by dynamically generating an effective $\mu$-parameter from the VEV of the scalar field, it also allows for an increase of the mass  of the lightest neutral CP-even Higgs \cite{NMSSMbound} (see \cite{Delgado:2010uj} 
for a version of the NMSSM which focuses on solving the little hierarchy problem, at the expense of not addressing the $\mu$-problem). In this paper we take a different approach by promoting the Higgs sector soft terms to supersymmetric operators, considering the degrees of freedom associated with the spontaneous breaking of supersymmetry and analyzing how they affect the Higgs sector. 

The MSSM provides an effective description of spontaneous supersymmetry breaking in terms of a set of soft terms which explicitly break supersymmetry \cite{soft}. While this description is a good approximation when the scale at which supersymmetry is spontaneously broken is high, it does not capture the dynamics and interactions of the degrees of freedom, beyond those of the SM and their superpartners, which are present in low scale models. The generic additional degrees of freedom correspond to the goldstino fermion which arises as a consequence of the spontaneous breaking of (global) supersymmetry. In the presence of gravity the spin 3/2 gravitino absorbs the spin 1/2 goldstino, which becomes its longitudinal components, and acquires a mass  $m_{3/2}= f/(\sqrt{3}M_{\mathrm{P}})$, where $M_{\mathrm{P}}$ is the Planck mass and $f$ is the order parameter of  supersymmetry breaking. 
 
In this paper  we are interested in the case where $\sqrt{f}$ is of the order of a few TeV, in which, due to the supersymmetric equivalence theorem \cite{cddfg}, the approximately massless gravitino (with $m_{3/2}\approx 10^{-3}-10^{-2}$ eV) can  be replaced by its goldstino components, gravitational effects can be neglected and supersymmetry can be treated as an approximately global symmetry. Moreover, we will have in mind a weakly coupled hidden sector which exhibits $F$-term breaking and gives rise to a chiral superfield $X=x+\sqrt{2}\theta \psi_X+\theta^2 F_X$ where $\psi_X$ becomes the goldstino at low energies, $x$ its complex scalar superpartner, the sgoldstino, and $F_X$ the auxiliary field which acquires a non-vanishing VEV that breaks supersymmetry. In contrast to the conventional way of parametrizing spontaneous supersymmetry breaking, where a background spurion field is introduced containing only a constant auxiliary component, the  goldstino multiplet $X$ contains propagating scalar and fermion degrees of freedom and a dynamical auxiliary field. The spurion description is a good approximation when supersymmetry is broken at a high scale but it does not describe the degrees of freedom present in the goldstino supermultiplet, which should be included in low scale models.

In contrast to the goldstino, the sgoldstino is not protected by any symmetry and therefore it generically acquires a mass $m_x<\sqrt{f}$ when heavy states in some weakly coupled hidden sector\footnote{The case when $m_x>\sqrt{f}$ is associated with strong coupling physics and will not be considered here.} are integrated out. Therefore, unlike the gravitino mass which is only sensitive to the scale $\sqrt{f}$, the sgoldstino mass is sensitive to the microscopic physics of the hidden sector. At energies above $m_x$ the spectrum is approximately supersymmetric and supersymmetry can be realized linearly.  At energies well below $m_x$ the sgoldstino can be integrated out and supersymmetry is realized non-linearly. In \cite{Komargodski:2009rz} it was suggested that this can be achieved by replacing $X$ by  $X_{\mathrm{NL}}$, which satisfies a constraint $X_{\mathrm{NL}}^2=0$ that effectively integrates out the sgoldstino component by replacing $x\to \psi_X \psi_X/(2 F_X)$ (see \cite{Antoniadis:2011xi} for a recent discussion on constrained goldstino and matter superfields).

 By integrating out some heavy states connecting the hidden and the visible sectors, supersymmetric effective operators are generated which couple fields of the visible sector and the goldstino multiplet, often described as a spurion field.  In order to account for the dynamics and interactions of the goldstino fermion, it was prescribed in \cite{Komargodski:2009rz} to replace the spurion by the non-linear superfield $X_{\mathrm{NL}}$.  This prescription was applied\footnote{See \cite{AlvarezGaume:2010rt} for applications of \cite{Komargodski:2009rz} to inflation.} in \cite{Antoniadis:2010hs} to the MSSM where it was shown that due to interactions between the auxiliary field $F_X$ and the MSSM Higgs fields, additional quartic Higgs couplings appear when $F_X$ is integrated out via its equation of motion. With these additional couplings it was shown possible to raise the tree level mass of the lightest Higgs particle in the large $\tan\beta$ region for low values of $\sqrt{f}$ (around 5 TeV). The fact that this prescription makes use of $X_{\mathrm{NL}}$ implies that the sgoldstino, or any effects due to it, is not considered. In this paper we study an effective model involving a weakly coupled hidden sector which gives rise to an sgoldstino. In order for the model to be valid at energies around and above $m_x$, we take into account the dynamics and interactions of the sgoldstino.

The effective model we consider has manifest and linearly realized supersymmetry and describes two Higgs doublets coupled to the dynamical goldstino superfield $X$ via operators that are supersymmetric realizations of the Higgs sector soft terms. In addition to a supersymmetric vacuum, the model exhibits meta-stable vacua\footnote{In the Conclusions we give a rough estimate of the life-time of these vacua and show that they can easily be sufficiently long-lived.} in which supersymmetry and electroweak (EW) symmetry are spontaneously broken. By including all the degrees of freedom corresponding to the goldstino multiplet we show that  additional couplings beyond those of the MSSM are induced and  when the supersymmetry breaking scale is low, these contributions significantly affects the Higgs sector. Since all effects and contributions beyond the usual ones in the MSSM arise from the same supersymmetric operators that give rise to the standard soft terms, the magnitude of these additional contributions is determined in terms of the parameters that determine the soft terms. The usual MSSM conditions for EW symmetry breaking are modified and the Higgs mass spectrum is obtained by considering fluctuations around these meta-stable vacua.    

We study in detail how the Higgs mass spectrum depends on the mass of the sgoldstino scalar.  The presence of the sgoldstino implies one extra CP-even and one extra CP-odd scalar state in the neutral Higgs sector. We discuss separately the two cases when the lightest Higgs particle is either doublet-like or sgoldstino-like, since they correspond to two different meta-stable vacua.  In the first (doublet-like) case, even though any mixing with the sgoldstino state reduces the mass of the lightest Higgs particle, it is possible to raise the  tree level masses for all the scalars above the LEP bound, for any value of $\tan\beta$. The second (light sgoldstino-like) case is richer and allows for more novel opportunities. First of all, it is possible to evade all present experimental mass bounds, already at tree level. Second, the spectrum contains a light CP-even and a CP-odd scalar particle. The presence of these light particles opens a region in the parameter space $(\mu, B_{\mu}, \tan \beta)$ where non-standard decays for doublet-like particles are possible, allowing for a richer phenomenology than in the corresponding MSSM parameter region. 

The outline of the paper is as follows: In Section 2 we present and discuss the model. In Section 3 the meta-stable vacua for the heavy sgoldstino case is considered, and the corresponding mass spectrum is discussed. In Section 4, the light sgoldstino case is analyzed, and the regions of large and small $\tan \beta$ are  studied separately. 
We will mainly focus on the first region, where the solutions are under better control and the mass spectrum is more interesting. Finally, we conclude in Section 5 and provide some details in the Appendices.

\section{The Low Energy Effective Model}
The  model we study has linearly realized supersymmetry and describes two Higgs doublets and a dynamical goldstino multiplet. It contains a Higgs sector and an effective hidden sector, along with operators that couple them, 
\begin{equation}
\label{L}
\mathcal{L}=\mathcal{L}_H+\mathcal{L}_X+\mathcal{L}_{H,X}~.
\end{equation}
The visible Higgs sector is simply
\begin{eqnarray}
\label{LH}
\mathcal{L}_H&=& \int d^4 \theta \, \sum_{i=1}^{2} H_{i}^{\dagger} e^{V} H_i 
+\left\{ \int d^2 \theta \,\mu H_1\cdot H_2+ \mathrm{h.c.} \right\}\nn \\
&&+\sum_{i=1}^{2}\left\{\frac{1}{16g_i^2}\int d^2\theta \left( 1-\frac{c_{\lambda_i}}{M}X \right)\Tr W_{(i)}^\alpha W_{(i)\alpha}+ \mathrm{h.c.} \right\}~
\end{eqnarray}
where $H_1\cdot H_2=H_{1}^{0} H_{2}^{0}-H_{1}^{-} H_{2}^{+}$ and $g_i$ are the gauge couplings for the $U(1)_Y$ and $SU(2)_L$ factors. The interactions of the $U(1)_Y$ and $SU(2)_L$ vector supermultiplets  with the Higgs supermultiplets are represented by the factor $e^V$, with $V$ in the appropriate representation. In our discussion, the only purpose of the gauge kinetic term\footnote{In the remainder of this paper, $g_1$ and $g_2$ will denote the normalized gauge couplings which take into account the correction due to the VEV of the scalar component of $X$.}, as well as the operator giving rise to the gaugino mass terms, will be to provide the $D$-term scalar potential. Since we have an explicit $\mu$-parameter in (\ref{LH}) we do not address the so called ``$\mu$-problem''. Instead, the idea is to consider a supersymmetric version of the MSSM soft terms for the Higgs sector and since the $\mu$-term is already a supersymmetric operator we simply leave it as it is.

The hidden sector is taken to be a simple Polonyi model,
\begin{equation}
\label{LX}
\mathcal{L}_X=\int d^4 \theta \, X^\dagger X\left( 1- \frac{c_X}{4M^2}X^\dagger X \right) +\left\{ \int d^2 \theta fX \,+\mathrm{h.c.} \right\}
\end{equation}
which provides a universal low-energy description of a weakly coupled hidden sector in which supersymmetry is spontaneously broken. The non-renormalizable operator in the Kahler potential generically arises as a consequence of integrating out some heavy states in the microscopic theory, e.g.~massive fields integrated out at 1-loop in an O'Raifeartaigh model or a massive vector field integrated out at tree level. This operator, to zeroth order in the VEVs of the scalars, gives a soft mass $m_x^2=c_X f^2/M^2$ to the sgoldstino which is sensitive to the microscopic physics in the hidden sector. However, in Section 4 we consider the case where $c_X=0$ and where the sgoldstino soft term only arises as a consequence of EW symmetry breaking.

The soft terms of the Higgs sector are assumed to arise from the following operators coupling the Higgs superfields and the goldstino superfield,
\begin{equation}
\label{LHX}
\mathcal{L}_{H,X}=\int d^4 \theta \, \sum_{i=1}^{2} - \frac{c_i}{M^2} X^\dagger X H_{i}^{\dagger} e^{V} H_i  -\left\{ \int d^2 \theta \, c_B X H_1\cdot H_2 +\mathrm{h.c.} \right\}~
\end{equation}
where the cutoff scale $M$ corresponds to the mass of some heavy states, connecting the visible and hidden sectors, that have been integrated out. Note that  the mass scale that suppress the non-renormalizable operator in (\ref{LX}) does not necessarily coincide with the one in (\ref{LHX}), even though we have for convenience denoted them in the same way. Out of many plausible operators, the ones appearing in (\ref{LHX}) are those with the lowest operator dimensions that, to zeroth order in the VEVs of the scalars,  give rise to the soft terms $m_i^2=c_i f^2/M^2$ and $B_\mu=c_B f$ in the Higgs sector. If $X$ in (\ref{LHX}) were treated as a spurion, these soft terms would be the only terms the Lagrangian (\ref{LHX}) gives rise to. Instead, since we are here treating all the components of $X$ dynamically, (\ref{LHX}) gives rise to interactions between all the components of the Higgs and the goldstino multiplets. 

Since $M$ denotes the cut-off scale of the effective model in (\ref{L}) we demand that all other scales are smaller than $M$. In particular, when the fields take VEVs, the  non-renormalizable terms in (\ref{L}) will correct the kinetic terms for the goldstino and Higgs fields. In order for these corrections to be under control, the VEV of the scalar component of any of the three chiral superfields is required be smaller than $M$. Also, in order to avoid higher-dimensional operators involving covariant superspace derivatives, the VEV of the auxiliary component of any of the chiral superfields is required to be smaller than $M^2$.  In terms of the Higgs sector parameters\footnote{Without loss of generality for the discussion in this paper, we take all the parameters of (\ref{L}) to be real.} we will demand that $\mu^2$,  $m_i^2$, $B_\mu$ and also $m_x^2$ are all smaller than $f$ in order for our effective model to be perturbative and reliable. 

Note that, from a strict effective field theory point of view, there are many other operators that could be included \cite{Brignole:2003cm,extraops}. However, our aim is not to consider the most general effective Lagrangian but instead to consider a supersymmetric realization of the MSSM Higgs sector, in which each soft term is promoted to the supersymmetric operator with lowest dimension that gives rise to that soft term. Since $M$ is not an independent parameter the only parameters beyond the ones already present in the MSSM are $f$ and $c_X$. In section 4 we will further restrict the model by considering the case where $c_X=0$. 

From the $F$-term equations of (\ref{L}) it is evident that there exists a supersymmetry preserving vacuum located at 
\begin{eqnarray}
\langle h_1\rangle_{\mathrm{susy}}\cdot \langle h_2 \rangle_{\mathrm{susy}}&=&\frac{f}{c_B}\label{v2susy} \\
 \langle x \rangle_{\mathrm{susy}}&=&\frac{\mu}{c_B} \label{vx2susy}~.
\end{eqnarray}
In the limit where $c_B\to 0$ the superpotential coupling in (\ref{LHX}) vanishes and the  supersymmetric vacuum (\ref{v2susy}) and (\ref{vx2susy}) is sent to infinity in all directions.

In Sections 3 and 4 we will be interested in finding supersymmetry breaking vacua in which the EW symmetry is broken in a phenomenologically viable way. Therefore, we will impose that  $\sqrt{f}$ is well above the EW breaking scale at $174$ GeV. Such vacua, due to the existence of the supersymmetric vacuum in (\ref{v2susy}) and (\ref{vx2susy}), is necessarily meta-stable. In order to guarantee that the vacua are indeed local minima we will demand that all the physical fluctuations have positive masses.

\subsection{The Scalar Potential}

\noindent The  $F$-term scalar potential obtained from (\ref{L}), together with the $D$-term gauge contributions, is up to quartic order in fields\footnote{The quartic order in fields will be sufficient for the analysis of the heavy sgoldstino case in Section 3 but in the light sgoldstino case in Section 4 we will need to consider higher orders.} given by 
\begin{eqnarray}
\label{V}
V  = V_{F}^{(f,x)}+ V_{F}^{(h^2)}+V_{F}^{(h^4)}+V_D
\end{eqnarray} 
where
\begin{eqnarray}
\label{Vfx}
V_{F}^{(f,x)} & = & f^2+c_X \frac{f^2}{M^2}| x|^2 + c_{X}^2 \frac{f^2}{M^4}| x|^4 \\
\label{VFh2}
V_{F}^{(h^2)} & = & \left( c_1 \frac{f^2}{M^2}+|\mu-c_B x|^2 +\left[ c_1 \frac{f^2}{M^4} \left( c_1+2 c_X \right)+ c_2 \frac{\mu^2}{M^2}\right] |x|^2 \right) |h_1 |^2  \nn \\
&&+\left( c_2 \frac{f^2}{M^2}+|\mu-c_B x|^2 +\left[ c_2 \frac{f^2}{M^4} \left( c_2+2 c_X \right)+ c_1 \frac{\mu^2}{M^2}\right] |x|^2 \right) |h_2 |^2  \\
&&+\left\{ \left( -c_B f+\frac{f}{M^2} \left[ (c_1+c_2)\,\mu \,\bar{x}-c_B(c_{1}+c_{2}+c_{X})|x|^2 \right] \right) h_1\cdot h_2 +\mathrm{h.c.}\right\}\nn\\
\label{VFh4}
V_{F}^{(h^4)} & = &\left| c_{1}\frac{f}{M^2}|h_1|^2+c_{2}\frac{f}{M^2}|h_2|^2-c_B\, h_1\cdot h_2 \right|^2\\
V_{D} & = &  \frac{g_{2}^{2}+g_{1}^{2}}{8} \left( |h_1|^2- |h_2|^2 \right)^2 + \frac{g_{2}^{2}}{2}|h_{1}^\dagger h_2|^2\label{VDterm}~.
\end{eqnarray}
By removing all couplings involving the sgoldstino $x$ we recover the MSSM scalar potential from (\ref{VFh2}) and (\ref{VDterm}). Note that the quartic Higgs couplings in (\ref{VFh4}) do not vanish in this limit since they arise as a consequence of integrating out the auxiliary component of the goldstino multiplet via its equation of motion. However, if this auxiliary component is not treated dynamically, as in the spurion approach, the couplings in (\ref{VFh4}) are not present.

\section{The Heavy Sgoldstino Case}

In this section and in the following one we  study supersymmetry and EW symmetry breaking meta-stable vacua of (\ref{V}). In such vacua the spectrum of physical scalar particles contains three CP-even neutral bosons and two CP-odd neutral bosons, including the complex sgoldstino, and a pair of charged bosons. In addition the spectrum contains three Goldstone bosons, one neutral and a pair of charged ones, which become the longitudinal components of the $Z$ and $W$ bosons.  We discuss separately the two cases where the smallest diagonal entry in the mass matrix for the CP-even Higgs bosons corresponds to either a Higgs doublet state or the sgoldstino state. In this section we focus on the heavy sgoldstino case. See Appendix A for a description of the scenario when the sgoldstino is integrated out and the scalar potential is written as a general doublet Higgs model. 

In order to look for meta-stable vacua of (\ref{V}) we give the following VEVs to the scalar fields
\begin{eqnarray}
\label{vevs}
\langle h_1 \rangle&=& \left(\begin{array}{c}\langle h_{1}^{0}\rangle \\ \langle h_{1}^{-}\rangle\end{array}\right)   =\left(\begin{array}{c}v_1 \\0\end{array}\right)=\left(\begin{array}{c}v \cos\beta \\0\end{array}\right)\nn\\
\langle h_2\rangle&=&\left(\begin{array}{c}\langle h_{2}^{+}\rangle \\ \langle h_{2}^{0}\rangle\end{array}\right)  =\left(\begin{array}{c}0 \\v_2\end{array}\right)=\left(\begin{array}{c}0 \\v\sin\beta\end{array}\right)\\
\langle x\rangle&=&v_x\nn~,
\end{eqnarray}
and impose that $v=174$ GeV$<<\sqrt{f}$. When minimizing the potential (\ref{V}) with respect to the sgoldstino we obtain the following expression for its VEV,
\begin{equation}
\label{EWx}
v_x= \frac{ \mu\, v^2\left( c_B-\frac{(c_1+c_2 )}{2}\frac{f}{M^2}\sin2\beta\right) }
{c_X \frac{f^2}{M^2} + v^2\Delta }
\end{equation}
where the order $v^2$ correction term in the denominator  is given by
\begin{eqnarray}
\Delta&=&  \left[ c_1 \frac{f^2}{M^4} \left( c_1+2 c_X \right) + c_B^2+c_2 \frac{\mu^2}{M^2}\right] \cos^2\beta \\
&&+ \left[ c_2 \frac{f^2}{M^4} \left( c_2+2 c_X \right)+ c_B^2+c_1 \frac{\mu^2}{M^2}\right] \sin^2\beta-c_B \frac{f}{M^2}  (c_{1}+c_{2}+c_{X}) \sin2\beta  ~.  \nn
\end{eqnarray}
Minimization with respect to the two Higgs scalars gives, up to order $v^2$ and $1/c_X$, the following two  conditions, 
\begin{eqnarray}
\label{EWh1}
\frac{2c_B}{\sin2\beta} +\frac{(c_1-c_2)}{\cos2\beta}\frac{f}{M^2}+ \frac{m_{Z}^{2}}{f} =\frac{v^2}{f}\left( c_B^2-\frac{2c_B \mu^2}{ f \sin2\beta} + \frac{4M^2 \mu^4}{c_X f^3}\left[ \frac{c_B}{\sin2\beta}-\frac{\mu^2}{f}  \right]\right) 
\end{eqnarray}
\begin{eqnarray}
\label{EWh2}
-\frac{2c_B}{\sin2\beta}+(c_1+c_2)\frac{f}{M^2}+\frac{2\mu^2}{f}&=&-\frac{v^2}{f}\Bigg(c_B^2+\frac{4\mu^4}{f^2}-\frac{2c_B\mu^2}{f}\left[ \sin2\beta +\frac{1}{\sin2\beta} \right]  \nn \\
&&\phantom{-\frac{v^2}{f}\,\,\,\,\,}+\frac{4c_B M^2 \mu^4}{c_X f^3}\left[ \frac{1}{\sin2\beta}  -\sin2\beta\right]  \Bigg)
\end{eqnarray}
where $m_Z^2=(g_1^2+g_2^2)v^2/2$ is the $Z$-boson mass. Imposing (\ref{EWh1}) and (\ref{EWh2}) on (\ref{EWx}) we get that the VEV of $x$ is up to order $v^2$ given by
\begin{equation}
\label{EWxinEW}
v_x= \frac{ \mu^3  v^2}{m_x^2f} \sin2\beta ~.
\end{equation}   
From (\ref{EWh1}) and (\ref{EWh2}) we see that once we solved for $c_1 f/M^2$ and $c_2 f/M^2$ the only dependence on $M$ appears in ratio with $c_X$. Therefore, we choose to express (\ref{EWx}) as well as the masses below in terms of the leading order (in VEVs) sgoldstino soft mass $m_{x}^2=c_X f^2/M^2$.

 Taking	into	account the kinetic normalization, the masses for the real and imaginary sgoldstino-like particle are, to order $v^2$ and $1/m_x^2$,  given by
\begin{eqnarray}
m_{\mathrm{Im}x}^2 & = & m_{x}^2
+v^2\Bigg[ \frac{m_{x}^2}{f}\left(2c_B  \sin2\beta-\frac{3\mu^2}{f}\right) +\frac{c_B\mu^2}{f}\left(  \frac{2}{\sin2\beta}-\sin2\beta \right) \nn \\
&&\phantom{c_X\frac{f^2}{M^2}+v^2~~}+ \frac{4\mu^2}{m_{x}^2}\left( \frac{\mu^2}{f}-\frac{c_B}{\sin2\beta} \right)^2 \Bigg] \label{Rexm}
\end{eqnarray}
\begin{eqnarray}
m_{\mathrm{Re}x}^2 & = &  m_{\mathrm{Im}x}^2+\frac{4 c_B \mu^2}{m_{x}^2}v^2\left[  \frac{2\mu^2}{f}\sin2\beta - c_B\right]\label{Imxm}~.
\end{eqnarray}
where $m_{\mathrm{Im}x}^2$ and $m_{\mathrm{Re}x}^2$ are eigenvalues of the 3$\times$3 scalar and pseudoscalar mass matrices, respectively, and differ from the sgoldstino soft mass $m_x^2$ by $v^2$-order corrections.  The masses for the other neutral Higgs particles are, up to order $v^2$ and $1/m_x^2$, given by
\begin{eqnarray}
m_{A^0}^2 & = & \frac{2 c_B f}{\sin2\beta} + v^2c_B\Bigg[ 2c_B-\frac{2\mu^2}{f\sin2\beta}+\frac{4\mu^2}{m_{x}^2\sin2\beta}\left( \frac{\mu^2}{f }- \frac{c_B}{\sin 2\beta}\right)\Bigg]
\end{eqnarray}
\begin{eqnarray}
m_{H^{0}}^2 & = &  m_{Z}^2 \sin^2 2\beta+m_{A^0}^2 +\frac{4c_B\mu^2}{m_{x}^2\sin2\beta}\,v^2\Bigg[ \frac{\mu^2}{f}- \frac{c_B}{\sin 2\beta}\Bigg]
\end{eqnarray}
\begin{eqnarray}
\label{mh}
m_{h^0}^{2} & = & m_{Z}^2 \cos^2 2\beta+v^2 \left[  \left(  \frac{ 2\mu^2}{f}-c_B \sin2\beta \right)^2 -\frac{4  \mu^6}{m_{x}^2 f^2}\sin^2 2\beta \right]~.
\end{eqnarray}
Here we have, with a slight abuse of notation, denoted the eigenvalues of the 3$\times$3 mass matrices using the standard MSSM notation even though they correspond to mass eigenstates mixed with the sgoldstino degrees of freedom. However, as seen in the expressions, since the soft mass of the sgoldstino $m_{x}^2$ is large, in this case such mixings are clearly small and vanishes in the limit where the sgoldstino effectively decouples. For all the choices of parameters  that we will consider (with $c_B>0$ in order to keep $m_{A^0}^2 $,$m_{H^{0}}^2>0$), the lightest physical particle will be $h^0$ and therefore we will focus on it.   

In the expression for $m_{h^0}^{2}$ in (\ref{mh}) we recognize the first term as the standard MSSM contribution arising from the $D$-term quartic Higgs couplings in (\ref{VDterm}). This contribution is represented in Figures 1a, 1b and 1d by the dashed curve. The second term in (\ref{mh}) is the contribution from the quartic Higgs couplings in (\ref{VFh4}) which arises as a consequence of integrating out the auxiliary component of the goldstino multiplet via its equation of motion and imposing the minimization conditions (\ref{EWh1}), (\ref{EWh2}) and (\ref{EWxinEW}). At large $\tan\beta$, the $c_B$-part of this term vanishes and the $\mu$-part dominates and provides a contribution that adds to the MSSM one, which is maximized in this regime.  
In the case where the sgoldstino is effectively decoupled, the large $\tan\beta$ region corresponds to the scenario discussed in \cite{Antoniadis:2010hs}, where it was shown that the $\mu$-part could raise the tree level mass above the LEP bound. At small $\tan\beta$, in contrast to the MSSM contribution which vanishes in this region, the $\tan\beta$-independent $\mu$-part contribution does not vanish. This is seen in Figure 1a where $m_{h^0}$ is given for different values of $\mu$ as function of $\tan\beta$, in the case where $c_B$ is small and the sgoldstino is  effectively decoupled. 


\begin{figure}[t!h!]
\def\baselinestretch{1.}
\begin{tabular}{cc|cr|}
\parbox{7.4cm}{
\subfloat[{\small $m_{h^0}$ as function of $\tan\beta$, varying $\mu$.}]
{\includegraphics[width=8.1cm]{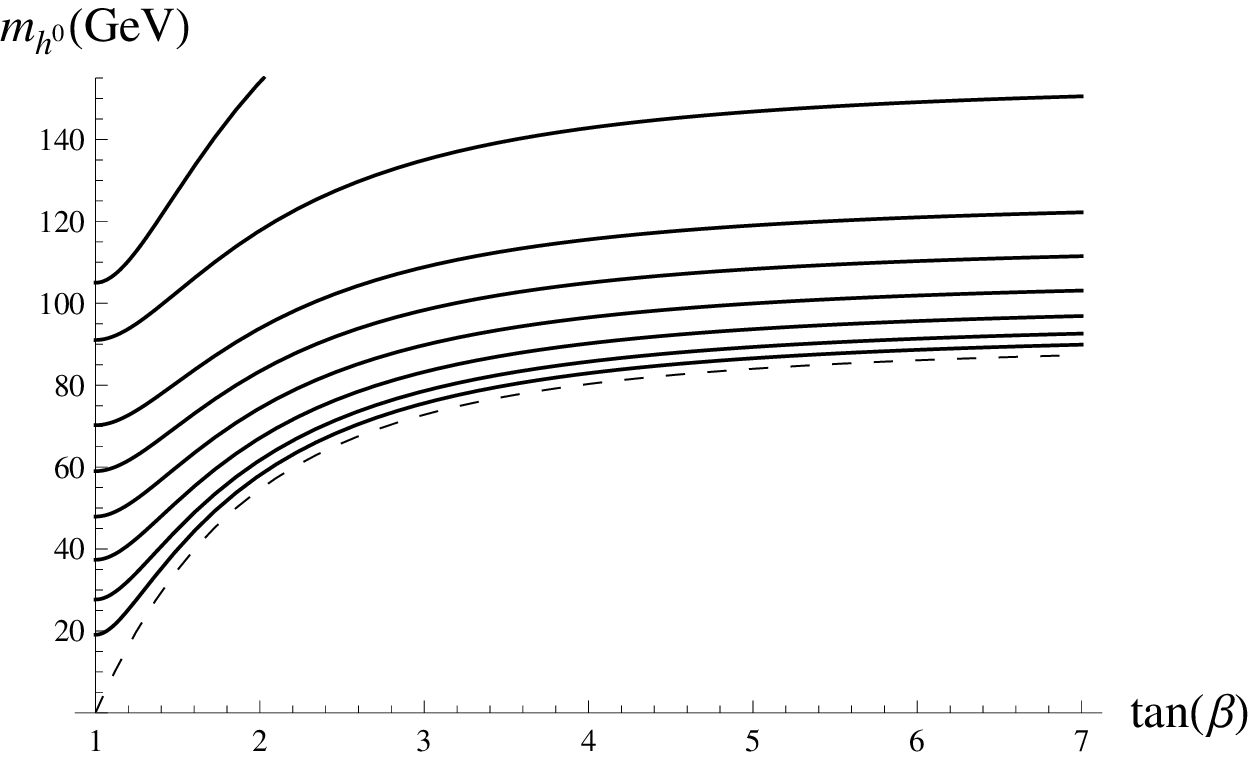}}}
\hspace{4mm}
\parbox{7.4cm}{
\subfloat[{\small $m_{h^0}$ as function of $\tan\beta$, varying $c_B$.}]
{\includegraphics[width=8.1cm]{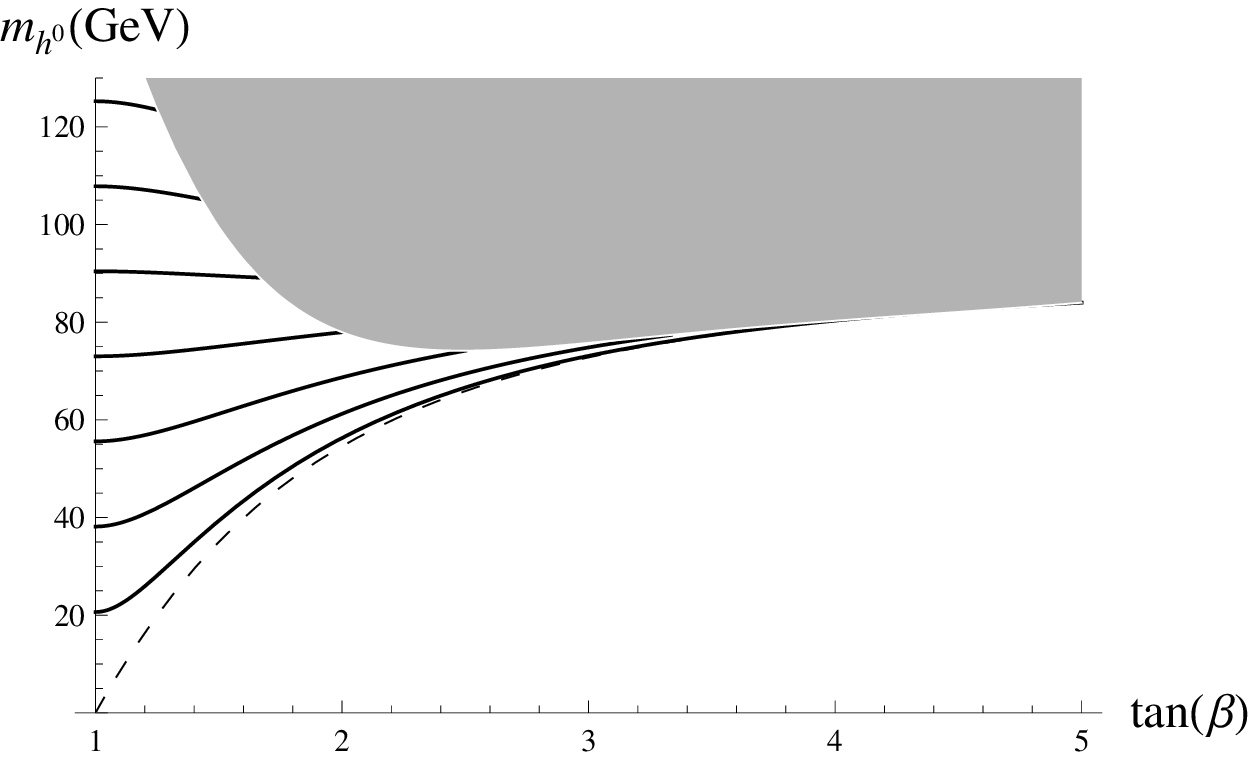}}}
\end{tabular}
\smallskip
\begin{tabular}{cc|cr|}
\parbox{7.5cm}{
\subfloat[{\small $m_{h^0}$ (in GeV) as function of $\mu$ and $c_B$.}]
{\includegraphics[width=8.1cm]{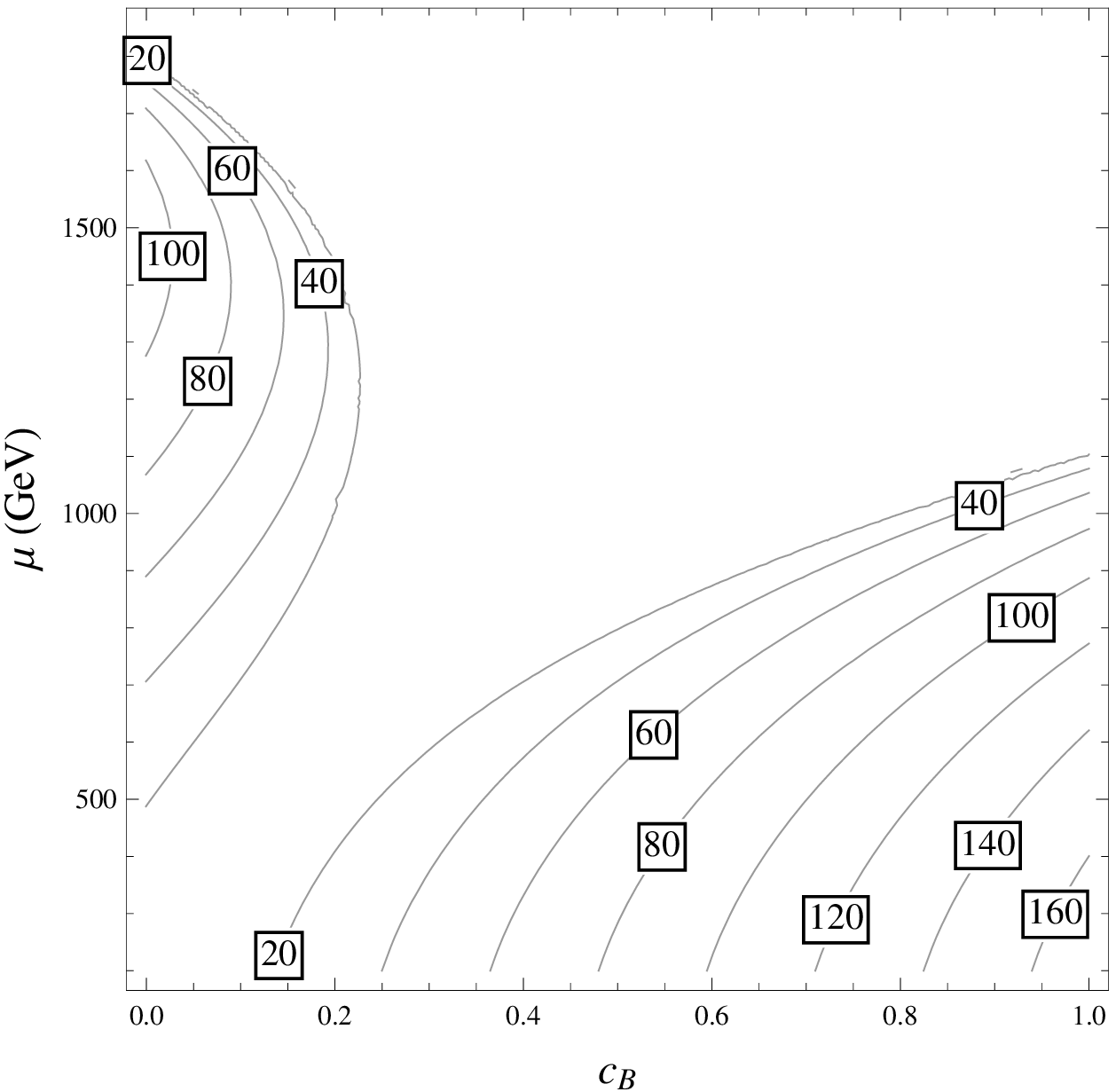}}}
\hspace{4mm}
\parbox{7.4cm}{
\subfloat[{\small $m_{h^0}$ as function of $\tan\beta$, varying $m_x$.}]
{\includegraphics[width=8.1cm]{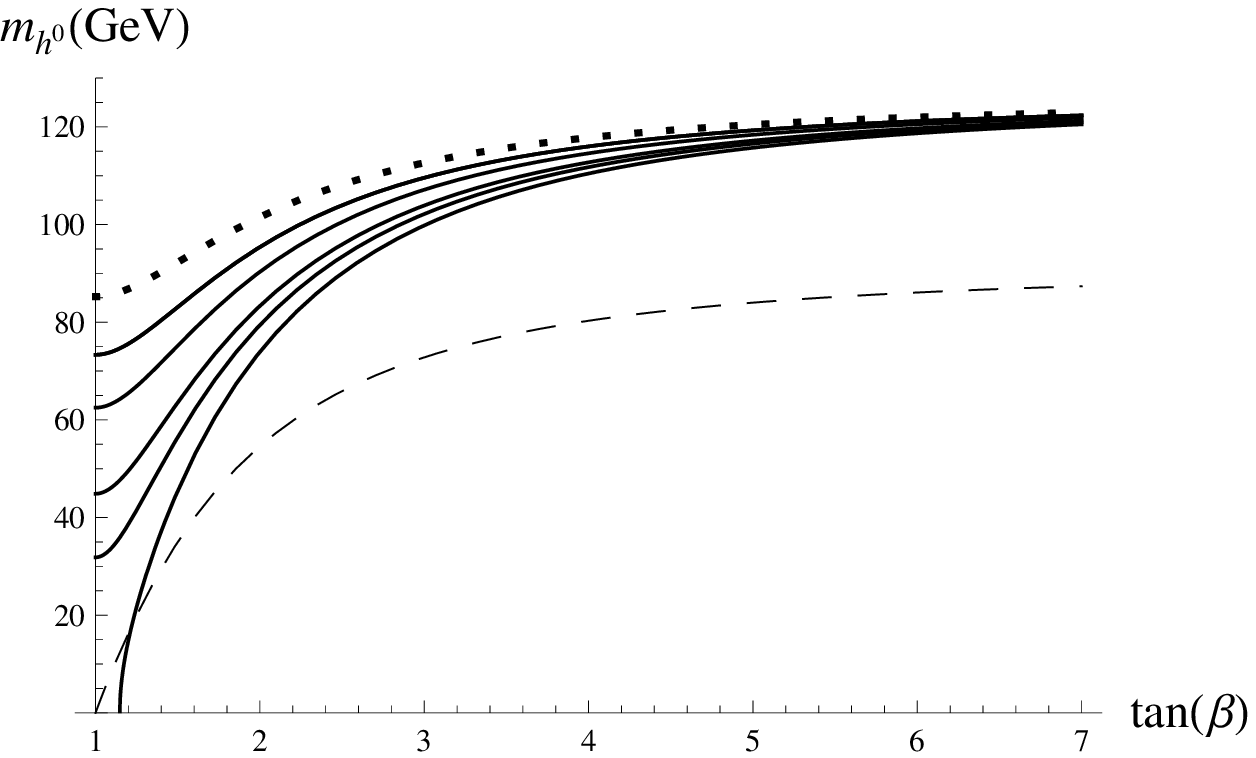}}}
\end{tabular}
\caption{{\protect\small
In these figures the tree level mass of the lightest CP-even neutral Higgs $m_{h^0}$ is given as a function of the different parameters. The dashed line corresponds to the MSSM value. In all plots we have fixed $\sqrt{f}=2$ TeV and in (a), (b) and (c) we have effectively decoupled the sgoldstino by setting $m_x=1.8$ TeV. In (a) $c_B=0.01$ and $\mu$ increases upwards from 500 GeV to 1000 GeV (in steps of 100 GeV) and with 1.2 TeV and 1.5 TeV corresponding to the two upper curves.  In (b) $\mu=400$ GeV and $c_B$ increases upwards from 0.2 to 0.8 (in steps of 0.1). The gray-shaded area corresponds to the region which is excluded by our assumptions on perturbativity.  In (c), $m_{h^0}$ is shown (in GeV) as a function of $\mu$ and $c_B$. In (d) $\mu=1$ TeV, $c_B=0.01$ and $m_x$ increases upwards from \{1, 1.1, 1.2, 1.5, 2 TeV\} to the dotted curve where the sgoldstino is completely decoupled. }}
\end{figure}


In the small $\tan\beta$ region, there is also the possibility of increasing the MSSM value for $m_{h^0}$ by considering the case where the $c_B$-part of the second term in (\ref{mh}) is dominating. This part gives rise to a contribution of the form $v^2c_B^2\sin^2 2\beta$, analogous to the contribution found in the context of the NMSSM, which is know to raise the lightest Higgs mass in the same region \cite{NMSSMbound}.  In Figure 1b, $m_{h^0}$ is given for different values of $c_B$ as function of $\tan\beta$, in the case where now $\mu$ is small. In the minimization conditions (\ref{EWh1}) and (\ref{EWh2}), when solving for $c_1f/M^2$ and $c_2f/M^2$, it can be seen that $c_1f/M^2$ scales with $\tan\beta$ as $c_1f/M^2\sim c_B \tan\beta$. Hence, for any value of $c_B$ there is an upper limit for $\tan\beta$ corresponding to the point where $c_1f/M^2$ becomes greater than one and invalidates our original assumptions. The excluded region is depicted by the  gray-shaded region. In Figure 1c we fix $\tan\beta=1$ and study the interplay between the $\mu$-part and the $c_B$-part of the second term in (\ref{mh}).

The third term in (\ref{mh}) corresponds to a contribution that arises due to mixing with the heavy sgoldstino-state. The fact that this contribution is negative is a consequence of the general fact that, if the off-diagonal entries can be treated perturbatively, the smallest eigenvalue of a mass matrix is bounded from above by the smallest diagonal entry. Due to the $\sin^2 2\beta$-dependence, this term vanishes in large $\tan\beta$ region, as can be seen in Figure 1d, where $m_{h^0}$  approaches the completely decoupled sgoldstino case, corresponding to the dotted curve. In Figure 2a and 2b we show the different dependence of $m_{h^0}$ on $\sqrt{f}$, for $\tan\beta=1$, in the two different regions of the parameter space corresponding to whether the $\mu$-part or $c_B$-part is dominating.

Let us stress that all the formulae for the masses are the tree level ones. The usual MSSM radiative corrections \cite{Higgs1loop} will also be present here and will for example further raise $m_{h^0}$.

\begin{figure}[t!h!]
\def\baselinestretch{1.}
\begin{tabular}{cc|cr|}
\parbox{7.4cm}{
\subfloat[{\small $m_{h^0}$ in function of $\sqrt{f}$, varying $\mu$}.]
{\includegraphics[width=8.1cm]{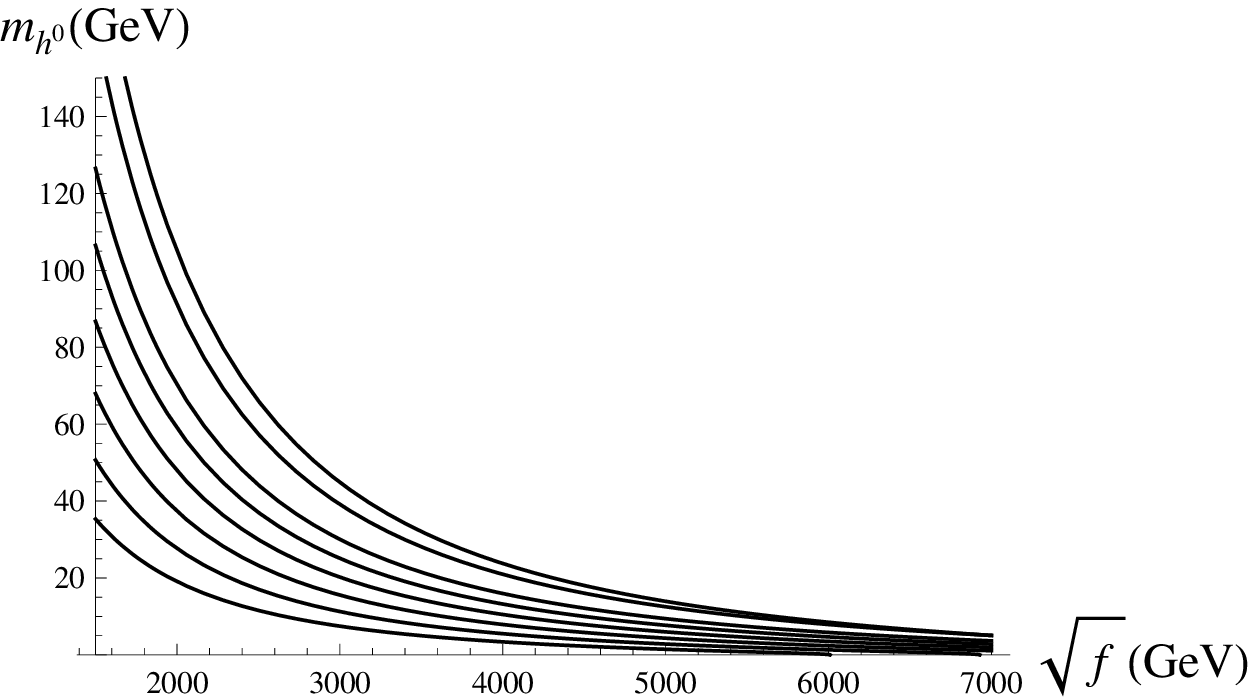}}}
\hspace{4mm}
\parbox{7.4cm}{
\subfloat[{\small $m_{h^0}$ in function of $\sqrt{f}$, varying $c_B$}.]
{\includegraphics[width=8.1cm]{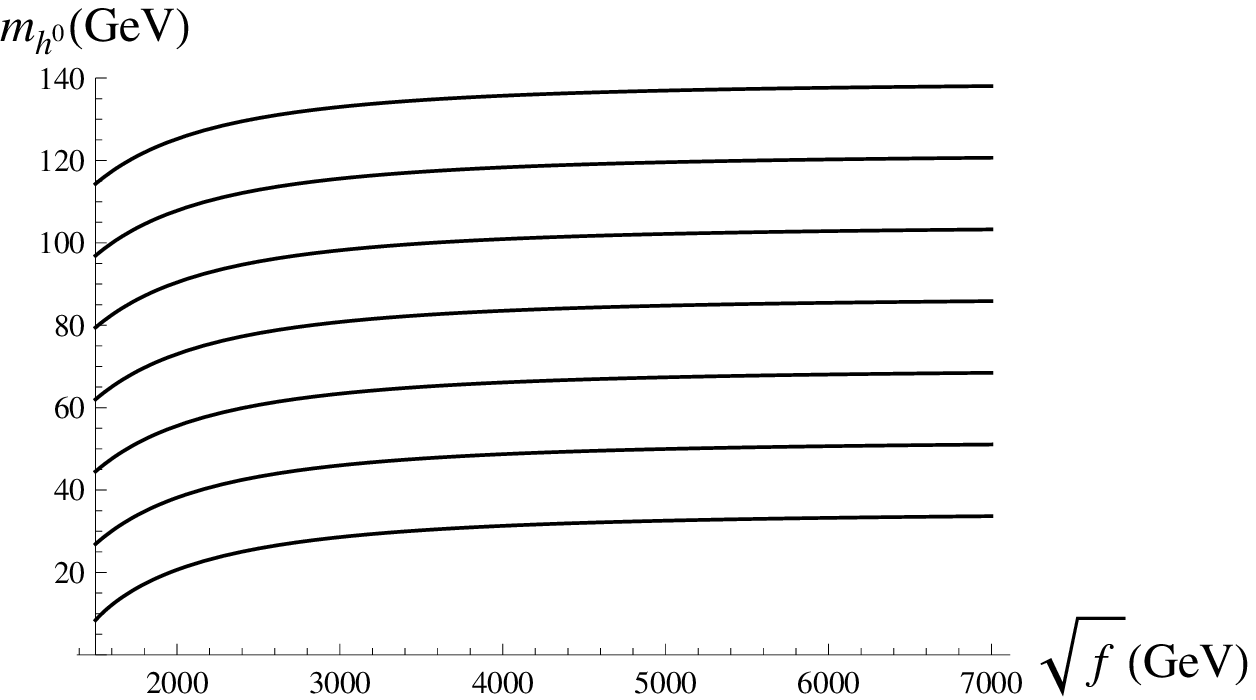}}}
\end{tabular}
\caption{{\protect\small
In these figures the tree level mass of the lightest CP-even neutral Higgs $m_{h^0}$ is plotted as a function of $\sqrt{f}$. In both figures we have fixed $\tan\beta=1$ and $m_x=1.8$ TeV. In (a) $c_B=0.01$ and $\mu$ increases upwards from 500 GeV to 1000 GeV (in steps of 100 GeV) and with 1.2 TeV and 1.5 TeV corresponding to the two upper curves. In (b) $\mu=400$ GeV and $c_B$ increases upwards from 0.2 to 0.8 (in steps of 0.1). }}
\end{figure}

\section{The Light Sgoldstino Case}

In this section we consider the case where the sgoldstino soft mass $m_x$ is taken to be smaller than $v=174$ GeV. In  this case, the VEV of the sgoldstino will generically be of order $v^0$. In order to take into account all possible hierarchies between the different parameters in the model, we separately consider the large and small $\tan \beta$ regions. Moreover, we will redefine the $c_X$ parameter in (\ref{LX}) as $c_X \rightarrow C_X \frac{v^2 M^2}{f^2}$, with $C_X$ being smaller than 1. In this way, the natural dynamically generated soft mass for the sgoldstino is $m_x^2 = C_X v^2$.
In both the large and small $\tan\beta$ regions, we consider also higher orders in fields in the scalar potential (\ref{V}), arising from the inverse of the Kahler metric and the equations of motion of the auxiliary $D$-fields of the vector supermultiplets. These additional contributions will be proportional to combinations of the VEV of the sgoldstino field and the couplings $c_1, c_2, C_X$, $g_1$ and $g_2$.  By requiring these corrections to be small and under control, we obtain useful criteria for excluding some solutions, and to put constraints on the parameters of the model.

\subsection{Large $\tan \beta $ }

Using the same strategy as in Section 3, we first look for solutions of the equations of motion for the different scalars, imposing the usual EW breaking VEVs for the Higgs fields. In this case, the VEV of $x$ and the EW symmetry breaking conditions for $c_1$ and $c_2$, will be expressed as expansions in of $C_X, v$ and $\gamma=(\tan \beta)^{-1}$. In particular, at  first order in $C_X$ and at second in $\gamma$ and $v$, we obtain the following  VEV for $x$
\begin{equation} \label{VEVxsmTinf}
v_x = \frac{f \gamma }{\mu } -  \gamma \left[\frac{f m_Z^2}{\mu ^3} + \frac{v^2 \mu }{f}\right]
- \frac{C_X f^3 \gamma }{2 \mu ^5}\left[1 -  \left(\frac{2 m_Z^2}{\mu ^2}+5 \frac{v^2 \mu ^2}{f^2} \right) \right]~.
\end{equation}
As expected,  the VEV of $x$ is generically different from the one obtained in the heavy sgoldstino scenario (\ref{EWxinEW}). Nevertheless both (\ref{EWxinEW}) and (\ref{VEVxsmTinf}) vanish in the limit $\tan \beta \rightarrow \infty$. 

In this large $\tan \beta$ regime, in analogy to the heavy sgoldstino case, the corresponding EW symmetry breaking conditions give rise  to a ``diverging" $c_1$ coefficient. In particular, the important ratio $\frac{c_1^2 f}{M^2} = \frac{m_1^2}{f}$ is proportional to $\frac{c_B}{\gamma}$. Since we demand perturbativity, $c_B$ is taken to be small compared to $\gamma$. Similarly to the redefinition of $c_X$, it is useful to keep track of this constraint by redefining $c_B \rightarrow C_B \gamma$, with $C_B $ being less than 1. In terms of these parameters, the EW symmetry breaking conditions read
\begin{eqnarray} \label{EWsmTinf}
&& \frac{c_1 f}{M^2} =   \frac{C_B}{2}\left[1 + \left( \frac{m_Z^2}{2 \mu ^2}+\frac{v^2 \mu ^2}{f^2}  \right) \right]  + c_{1,x} C_X + c_{1,\gamma} \gamma^2 \nonumber \\
&&  \frac{c_2 f}{M^2} = -\frac{\mu ^2}{f} \left[1+  \left( \frac{m_Z^2}{2 \mu ^2}+\frac{2 v^2 \mu ^2}{f^2}  \right) \right] + c_{2,x} C_X + c_{2,\gamma} \gamma^2
\end{eqnarray}
where the corrections $c_{1,x}, c_{2,x}, c_{1,\gamma}, c_{2,\gamma}$ are given in the Appendix B.1.

Concerning the reliability of these vacuum solutions, we have to be sure that the iterative procedure of solving the equations of motion with $C_X$ as a perturbative parameter is consistent with the final results. Since there are many scales in the model, this is not a priori obvious and the corrections to the leading order term could be the dominant ones. In order to avoid this problem, we see from (\ref{VEVxsmTinf}) that a safe choice at this order is to restrict the analysis to the region in parameter space where $|C_X| < 2 \frac{\mu^4}{f^2}$ (we always assume $v^2 < \mu^2 < f$). Moreover, a milder limitation is obtained by demanding that the ratios $\left| \frac{c_i f}{M^2} \right|$ and $\left| \frac{c_i v_x^2}{M^2} \right|$, with $i=(1,2)$, are less than one. In fact, the only non-trivial one corresponds to the constrained region $|C_B| < 2 \frac{\mu^2}{f \gamma^2}$, which is easily realized for a small enough $\gamma$, at any given $\mu$ and $f$. It should be stressed that the constraint we put on $C_X$ is not "physical", but is only related to our approximations. In principle, $C_X$ has only to be smaller than 1, but in that case, a different strategy should be used in order to find the vacuum of the model. 

Proceeding as in the previous section, after transforming the fields in order to have diagonal and canonically normalized kinetic terms, we obtain the mass matrices for the neutral scalars. The resulting mass eigenvalues are the following\footnote{We denote here $m_{\mathrm{Re}x}^2$ and $m_{\mathrm{Im}x}^2$ as the mass eigenvalues of the states which are dominantly sgoldstino-like and the others according to the usual MSSM notation with $h^0$ being the lightest neutral CP-even scalar coming from the Higgs doublets.}
\begin{eqnarray} \label{MsmTinf}
&& m_{\mathrm{Re}x}^2 =  m_{\mathrm{Im}x}^2 = \frac{4 C_B v^2 \mu ^4}{C_B f^2+2 f \mu ^2}   + \delta_{\mathrm{Re}x, C_X} ~\frac{m_{x}^2}{v^2}  + \delta_{\mathrm{Re}x, \gamma} ~\gamma^2\\
&& \nonumber \\
&& m_{H^0}^2 = \frac{C_B f}{2} \left(1 + \frac{m_Z^2}{2\mu^2} \right) 
+\mu^2+ v^2 \left(C_B^2 -\frac{m_Z^2}{2 v^2}+\frac{\mu ^4}{f^2}
-\frac{C_B^3 f }{C_B f+2 \mu ^2}
- \frac{2 C_B 
\mu ^2}{f} \right)\nonumber \\
&& \qquad \qquad + \delta_{H^0, C_X} ~\frac{m_{x}^2}{v^2}   + \delta_{H^0, \gamma}~ \gamma^2 \\
&& \nonumber \\
&& m_{A^0}^2 = m_{H^0}^2 + \Delta_A \gamma^2\\
&& \nonumber \\
&& m_{h^0}^2 = m_Z^2+\frac{4 v^2 \mu ^4}{f^2} -\gamma^2 v^2 \left(16~ C_B \frac{\mu^2}{f}+3~ \frac{m_Z^2}{v^2}+12 ~ \frac{\mu^4}{f^2}\right)
\end{eqnarray}
where the $\delta$'s and $\Delta_A$ corrections are given in the Appendix B.1. Note that $m_{h^0}^2$ is, at this order in $C_X$ and $\gamma$, independent of $C_X$. Thus, we can focus on the case $C_X=0$. The fact that the model can provide a good solution even without any  stabilizing term for the sgoldstino in the Kahler potential is interesting by itself, as we will discuss below.

\begin{figure}[t!h!]
\def\baselinestretch{1.}
\begin{tabular}{cc|cr|}
\parbox{7.4cm}{
\subfloat[{\small $m_{h^0}$ and $m_{\mathrm{Re}x}$  varying $\mu$.}]
{\includegraphics[width=8.2cm]{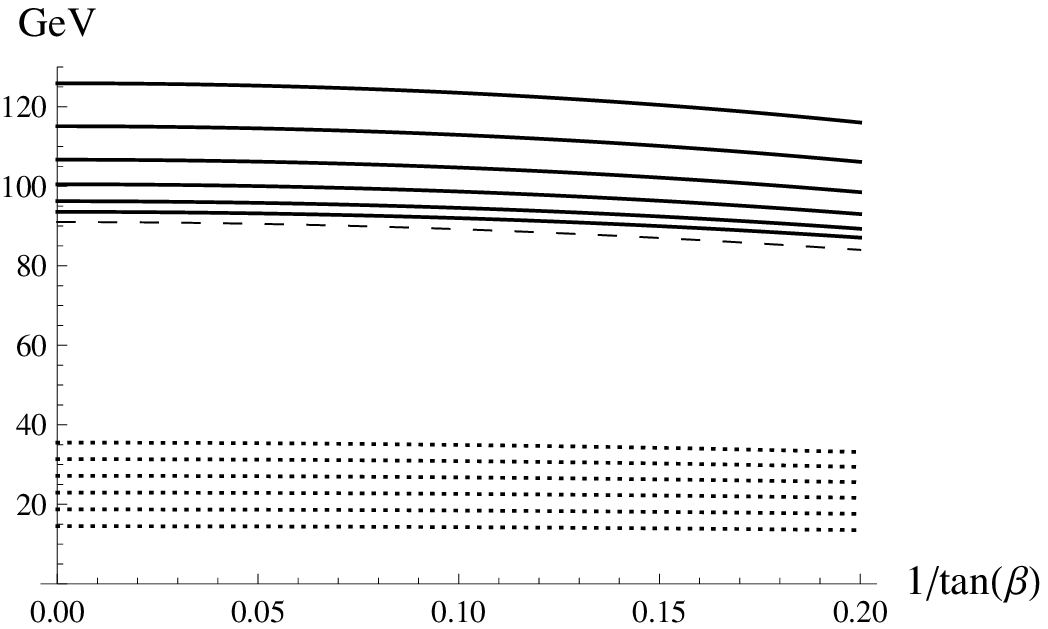}}}
\hspace{4mm}
\parbox{7.4cm}{
\subfloat[{\small $m_{h^0}$ and $m_{\mathrm{Re}x}$ varying $C_B$.}]
{\includegraphics[width=8.2cm]{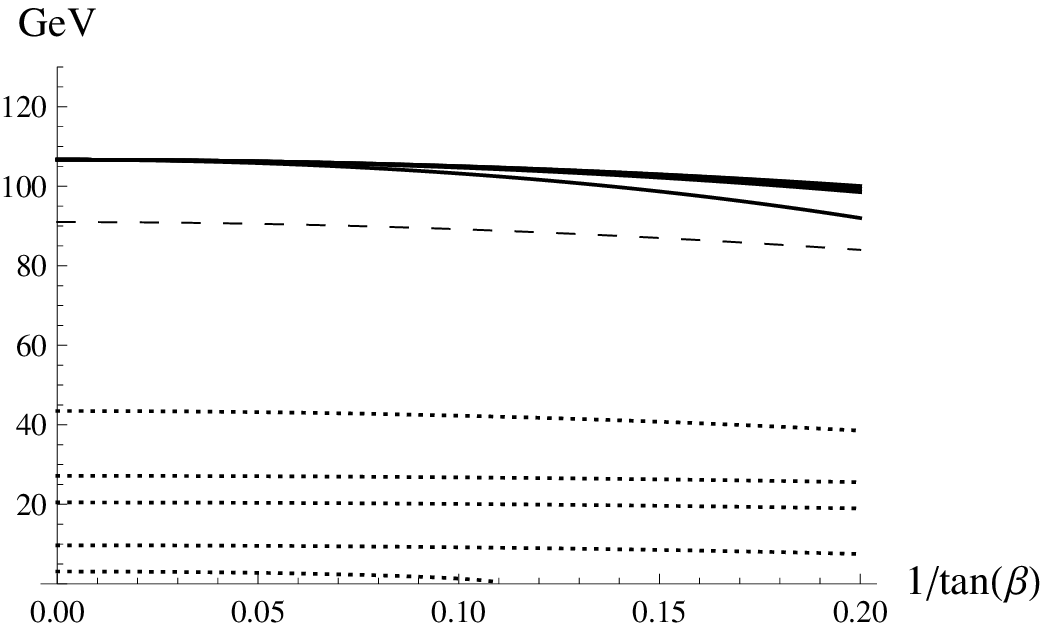}}}
\end{tabular}
\smallskip
\begin{tabular}{cc|cr|}
\parbox{7.5cm}{
\subfloat[{\small $m_{H^0}$ varying $\mu$.}]
{\includegraphics[width=8.2cm]{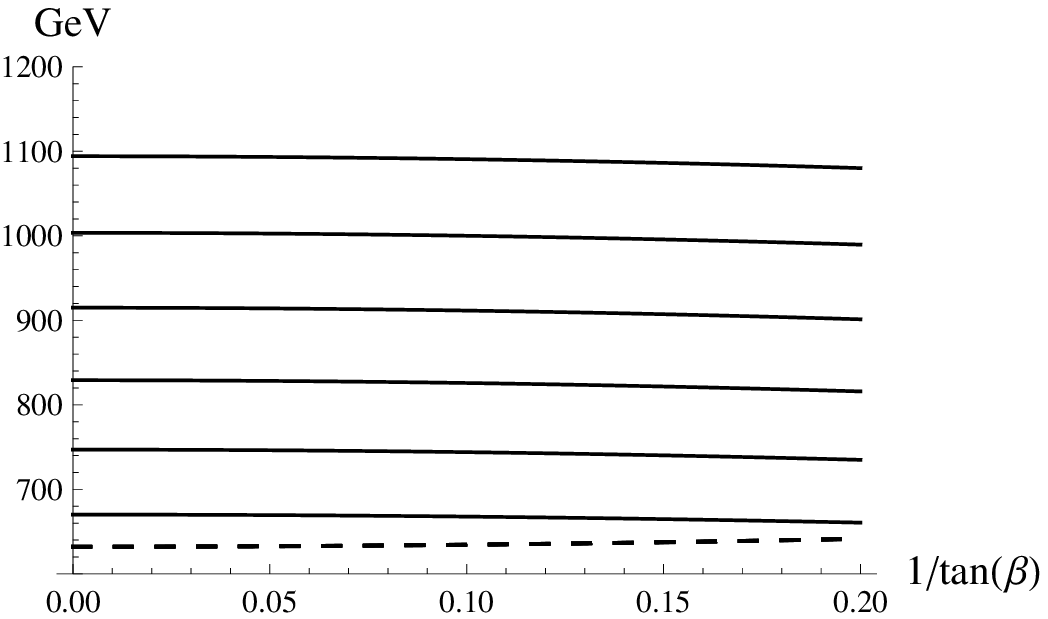}}}
\hspace{4mm}
\parbox{7.4cm}{
\subfloat[{\small $m_{H^0}$ varying $C_B$.}]
{\includegraphics[width=8.2cm]{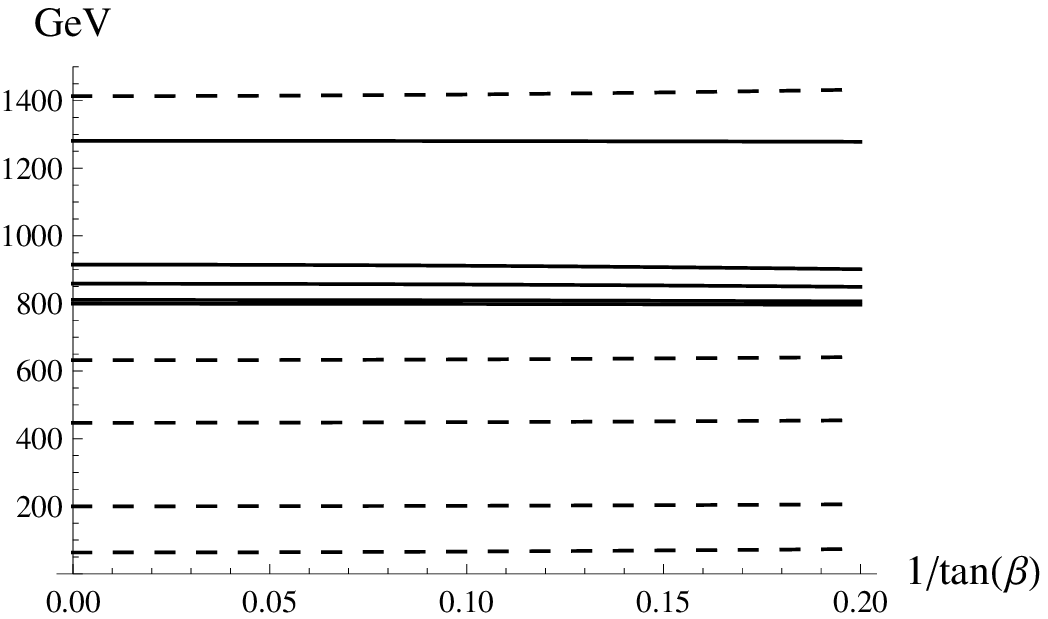}}}
\end{tabular}
\caption{{\protect\small
In these figures the tree level mass of the CP-even neutral scalars are given as functions of $(\tan \beta)^{-1}$, for different values of $\mu$ and $C_B$, in the case $C_X=0$. In (a) and (c) $C_B = 0.1$ and $\mu$ increases upwards from 500 GeV to 1000 GeV (in steps of 100 GeV). In (b) and (d) $\mu=800$ GeV and $C_B$ takes values (0.001, 0.01, 0.05, 0.1, 0.5). Again, in all plots we have fixed $\sqrt{f}=2$ TeV.  
In (a) and (b) the solid lines correspond to $m_{h^0}$, whereas the dotted ones to $m_{\mathrm{Re}x}$. With a dashed line we show the corresponding value for $m_{h^0}$ in MSSM. In (c) and (d) the solid lines represent $m_{H^0}$ in the light sgoldstino case, whereas the dashed ones correspond to $m_{H^0}$ for $m_x \rightarrow \infty$, with the same set of parameters.
 }}
\label{smallcx}
\end{figure}

As expected, the lightest particles arise in this case from the sgoldstino complex scalar, and they are mass degenerate at this order in the parameters. The next-to-lightest particle $h^0$, receives new contributions with respect to the MSSM case, in a  similar way to that shown in the previous section. In particular, we can see in Figure \ref{smallcx}a that both the tree-level values of $m_{h^0}$ and $m_{\mathrm{Re}x}$ increase with $\mu$ and already for a ratio $\frac{\mu^2}{f} \sim 0.05$ the value of $m_{h^0}$ stays above the MSSM one. At the same time, for a fixed value of $\mu$, varying the parameter $C_B$ affects  $m_{h^0}$ very mildly, whereas it allows for different values of the sgoldstino-like scalar mass, as is shown in  Figure \ref{smallcx}b. 

This spectrum of possibilities can be interesting in the context of experimental bounds on the Higgs scalar masses. First of all, as a general result in this large $\tan \beta$ regime, the lightest eigenstates are mostly sgoldstino-like, in the sense that they mix in a mild way with the Higgs doublet scalars. In fact, at  zeroth order in $\frac{m_x^2}{v^2}$ and $\gamma$ and, for example, at first order in $C_B$, one can show that the lightest mass eigenstates, in terms of the gauge ones, are given by
\begin{eqnarray}
&& \phi_{\text{lightest,CP-even}} =   \left[ 1 + \frac{v^2 }{f}\left( C_B - \frac{\mu^2}{2 f} \right) \right] \mathrm{Re} x - \frac{v}{\mu} \left( C_B - \frac{\mu^2}{ f} \right) \mathrm{Re}h_1 \nonumber \\
&& \phi_{\text{lightest,CP-odd}} = \left[ 1 + \frac{v^2 }{f}\left( C_B - \frac{\mu^2}{2 f} \right) \right] \mathrm{Im} x - \frac{v}{\mu} \left( C_B - \frac{\mu^2}{ f} \right) \mathrm{Im}h_1 ~.
\end{eqnarray}
In this regime, the mixing only involves $h_1$ (the mixing with $h_2$ is proportional to $1/\tan \beta$), which is the field that is mainly contributing to the heavy $H^0$ eigenstate. As long as $\mu$ stays relatively big compared to $v$, the mixing between the gauge eigenstates can  be small enough to evade the experimental bounds \cite{Schael:2006cr}. As discussed above, the same choice of parameters allows us to increase the tree-level mass of the next-to-lightest  CP-even particle $h^0$. 

Note that, in contrast to the heavy sgoldstino scenario, the $\mu$ and $C_B$ parameters can safely span a larger region in the parameter space. In fact, in Figure  \ref{smallcx}c, we show that for a given value of $C_B$, the mass of heavy doublet-like state $H^{0}$ increases with $\mu$ and in  Figure  \ref{smallcx}d that it is bounded from below (for a fixed $\mu$) once we decrease $C_B$. This is in contrast to the heavy sgoldstino case where the mass of $H^{0}$ was quite insensitive to changes in $\mu$ but proportional to $C_B$, making the small $C_B$ region delicate. This is mainly due to the fact that in the scenario discussed in the previous section, the dominant contribution to the $H^{0}$ mass is given by $C_B f$, as in the MSSM (which, in standard notation, reads $2 B_\mu /\sin 2\beta$), whereas in this case it is a combination of $C_B$- and $\mu$-depending contributions.

\begin{figure}[h!t!]
\begin{center}
\def\baselinestretch{1.}
{\includegraphics[width=8.4cm]{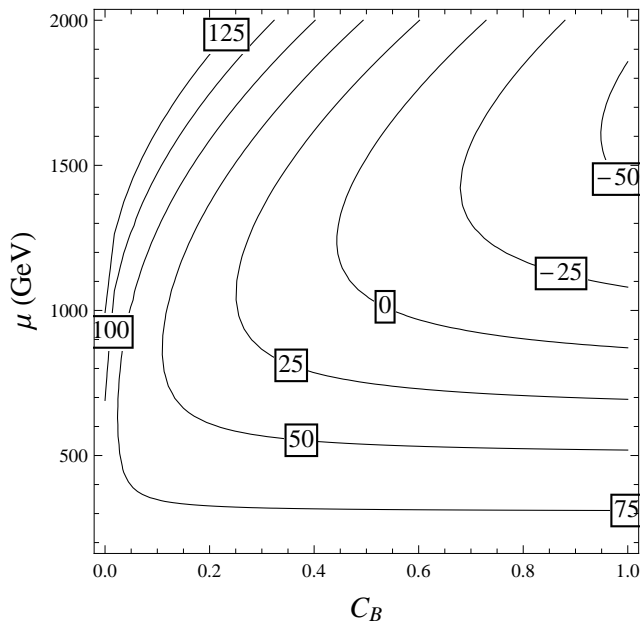}}
\end{center}
\caption{{\protect\small In this figure we show the difference in GeV between $m_{h^0}$ and twice the mass of the CP-even (or, equivalently, the CP-odd) sgoldstino scalar, in the limit $\tan \beta \rightarrow \infty$, fixing $\sqrt{f} = 2$ TeV.
 }}
\label{smallcxContour}
\end{figure}

The presence of extra light singlet scalars can  be important for non-standard Higgs decays, similarly to the discussion in \cite{Dermisek:2005ar} for the NMSSM. In fact, as shown in Figure \ref{smallcxContour}, by varying $\mu$ and $C_B$, the mass of the lightest particles can pass from a scenario where the decay of $h^{0}$ into two sgoldstino-like particles is forbidden  to one where it is kinematically allowed. However, a full analysis of these decays and the  related phenomenology\footnote{See \cite{sGoldphen} for discussions concerning different aspects of sgoldstino phenomenology.} goes beyond the scope of this paper.

\subsection{Small $\tan \beta $ }

In this regime, the analysis of the vacuum structure is more complicated and, at least in our perturbative approach, the solutions we obtain are less reliable. By defining $\tilde{\gamma} = \frac{1}{2} (\tan \beta -1)$ and considering the first order in $C_X$ and $\tilde{\gamma}$ and the second order in $v$, we find the following solution for $x$, 
\begin{eqnarray} \label{VEVxsmT1}
v_x =  \frac{\mu}{c_B} -\frac{C_X f}{c_B^2 \mu} \left( 1+\frac{3 c_B v^2}{f} \right) 
\end{eqnarray}
and the corresponding EW symmetry breaking conditions read
\begin{eqnarray} \label{EWsmT1}
&& \frac{c_1 f}{M^2} = \frac{c_B^2 f}{ \mu^2+c_B f} -\frac{c_B^4 v^2 f}{2 (\mu^2+c_B f)^2}+  \tilde{c}_{1,x} C_X + \tilde{c}_{1,\tilde{\gamma}} \tilde{\gamma} \nonumber \\
&& \frac{c_2 f}{M^2} =  \frac{c_B^2 f}{ \mu^2+c_B f} -\frac{c_B^4 v^2 f}{2 (\mu^2+c_B f)^2}+ \tilde{c}_{2,x} C_X + \tilde{c}_{2,\tilde{\gamma}} \tilde{\gamma}
\end{eqnarray}
where once again, the corrections  $\tilde{c}_{1,x}, \tilde{c}_{2,x}, \tilde{c}_{1,\tilde{\gamma}}, \tilde{c}_{2,\tilde{\gamma}}$ are given in the Appendix B.2.

Using the same arguments as before, our solutions are reliable only when $|C_X| < \left| \frac{c_B \mu^2}{f} \right|$. Moreover, imposing $\left|\frac{c_1 v_x^2}{M^2} \right| <1$ corresponds to demanding  $\left| \frac{\mu^2}{\mu^2+c_B f} \right|<1$.\\
In this vacuum, the mass eigenvalues are given by
\begin{eqnarray} \label{MsmT1}
&& m_{\mathrm{Re}x}^2 =  \frac{c_B v^2 \mu^2}{f}  +  \tilde{\delta}_{\mathrm{Re}x, C_X}  ~\frac{m_{x}^2}{v^2}\nonumber \\
&& m_{\mathrm{Im}x}^2 = - \frac{c_B v^2 \mu^2}{f}    + \tilde{\delta}_{\mathrm{Im}x, C_X} ~\frac{m_{x}^2}{v^2} \nonumber \\
&& m_{H^0}^2 = \frac{c_B^3 f v^2+c_B^2 (2 f^2+v^2 \mu^2)+c_B f (m_Z^2+4 \mu^2)+2 \mu^4}{c_B f+\mu^2} +  \tilde{\delta}_{H^0, C_X} ~\frac{m_{x}^2}{v^2} \nonumber \\
&& m_{A^0}^2 = 2 c_B^2 v^2+\frac{9 c_B v^2 \mu^2}{f}+2 c_B f+2 \mu^2 + \tilde{\delta}_{A^0, C_X}~\frac{m_{x}^2}{v^2} \nonumber \\
&& m_{h^0}^2 =  c_B^2 v^2+ \frac{c_B v^2 \mu^2}{f} +  \tilde{\delta}_{h^0, C_X} ~\frac{m_{x}^2}{v^2} ~.
\end{eqnarray}
The first corrections in $\tilde{\gamma}$ arise at order $\tilde{\gamma}^2$ (and $C_X \tilde{\gamma}$) and are therefore not taken into account. As  is evident, the masses of the CP-even and CP-odd sgoldstino-like scalars have different signs at the leading order. Moreover, as is shown in Appendix B.2, $\tilde{\delta}_{\mathrm{Re}x, x}$ and $\tilde{\delta}_{\mathrm{Im}x, x}$ are proportional to $v^2$. Once one imposes the constraint on $C_X = \frac{m_{x}^2}{v^2}$ discussed above, it is easy to see that at this order it is impossible to avoid  tachyonic directions without going beyond the reach of our approximations. Again, this discussion is not sufficient to conclude that there is no supersymmetry breaking vacuum in the region of small $\tan \beta$ for a light sgoldstino, but it excludes the possibility of obtaining a viable solution with $C_X$ treated in this perturbative way.

\section{Conclusions}

In this paper we discussed an effective model with manifest supersymmetry describing two Higgs doublets and a dynamical goldstino superfield $X$. The model corresponds to a supersymmetric realization of the MSSM Higgs sector with two additional terms in the Lagrangian, one being the linear Polonyi term in the superpotential, introducing the supersymmetry breaking scale $f$, and the other one being a quartic term for $X$ in the Kahler potential. The latter term provides a soft mass $m_x$ to the scalar component of $X$, the sgoldstino, which generically stabilizes the sgoldstino VEV. $X$ is coupled to the Higgs fields by promoting the soft terms to supersymmetric operators. The model allows for both a supersymmetric and meta-stable vacua, with characteristics depending on the relative hierarchy between $\sqrt{f}$,  $m_x$ and the scale of EW symmetry breaking $v=174$ GeV (where we always assume $\sqrt{f}>v$). The case where $m_x \geqslant \sqrt{f}$ corresponds to the non-linear realization of the MSSM discussed in \cite{Komargodski:2009rz,Antoniadis:2010hs}. In this paper we restrict ourselves to the perturbative scenario where $m_x < \sqrt{f}$. 

In the case where $m_x >v$ we found a meta-stable vacuum in which all the Higgs particles can be massive enough to evade the LEP lower mass bound, already at tree level. In the case where $m_x <v$ we found a different meta-stable vacuum in the large $\tan\beta$ region where the lightest CP-even and CP-odd particles are dominantly sgoldstino-like.  In this case, the scenario with $m_x=0$ is viable and particularly attractive since, for any choice of the supersymmetry breaking scale $\sqrt{f}$, it reduces the set of parameters of the model to the standard set $(\mu, B_{\mu}, \tan \beta)$ of the MSSM Higgs sector but being associated with non-standard phenomenology. In particular, this scenario allows for novel decays of doublet-like states into sgoldstino-like states while at the same time keeping the tree level masses of the doublet-like states significantly above the MSSM values. As a future direction we plan to include the other MSSM fields and perform a detailed analysis of the related phenomenology.   

Concerning the life-time of these meta-stable vacua we can give a rough estimate by evaluating the bounce action. For the meta-stable vacua we have considered, the bounce action is  $S_{\mathrm{B}}\approx\frac{(\langle x\rangle_{\mathrm{susy}} -v_x)^4}{f^2}\approx \mu^4f^2/B_{\mu}^4$, implying a sufficiently long life-time for a  small enough $B_{\mu}/f$. This is  easily achieved  in the large $\tan\beta$ regime where  perturbativity and EW breaking conditions generically require $B_{\mu}/f$ to scale as $1/\tan\beta$. On the other hand, in the parameter region where $B_{\mu}/f$ is not small this estimate is not sufficient to exclude the validity of the corresponding meta-stable vacua since an embedding  into a microscopic model can for example introduce new directions in field space contributing to the bounce action. In order for the picture to be more complete, in particular at the quantum level, it would be necessary to find a ultraviolet completion of this effective  model.

\begin{center}
{\bf Acknowledgements}
\end{center}
We thank E.\,Dudas, J.R.\,Espinosa, G.\,Ferretti, D.~M.~Ghilencea, P.~Tziveloglou for interesting discussions. This work was supported by the Spanish MICINNÕs Juan de la Cierva  and Consolider-Ingenio 2010 programme under grants CPAN CSD2007- 00042, FPA2009-07908, FPA2010-17747, MultiDark CSD2009-00064, the Community of Madrid under grant HEPHACOS S2009/ESP-1473 and the European Union under the Marie Curie-ITN programme PITN-GA-2009-237920.

\appendix

\section*{Appendices}

\section{Integrating Out the Sgoldstino}

\noindent In this appendix we consider the case where  sgoldstino is heavy and we integrate it out via its equation of motion. The effective model we obtain is valid at energies well below the sgoldstino mass and can be written in terms of a general two Higgs doublet potential
\begin{eqnarray}
\label{Vtot}
V_{\mathrm{tot}} & = & \widetilde{m}_{1}^{2} |h_1 |^2 +\widetilde{m}_{2}^{2}  |h_2 |^2 - \left( m_3^2 h_1\cdot h_2 + \mathrm{h.c.} \right) \nn \\
&& + \frac{\lambda_1}{2} |h_1 |^4 +\frac{\lambda_2}{2} |h_2 |^4 + \lambda_3
 |h_1 |^2  |h_2 |^2 + \lambda_4 |h_1\cdot h_2|^2 \nn \\
 &&+\left(\frac{\lambda_5}{2}(h_1\cdot h_2)^2+  \lambda_6 |h_1 |^2 h_1\cdot h_2 + \lambda_7 |h_2 |^2 h_1\cdot h_2 +\mathrm{h.c.} \right) ~
\end{eqnarray} 
where  
\begin{eqnarray}
\label{m2}
\widetilde{m}_{1}^{2} & = & c_1 \frac{f^2}{M^2}+(\mu-c_B v_x)^2 +\left[ c_1 \frac{f^2}{M^4} \left( c_1+2 c_X \right)+ c_2 \frac{\mu^2}{M^2}\right] v_x^2 \nn \\
\widetilde{m}_{2}^{2} & = & c_2 \frac{f^2}{M^2}+(\mu-c_B v_x)^2 +\left[ c_2 \frac{f^2}{M^4} \left( c_2+2 c_X \right)+ c_1 \frac{\mu^2}{M^2}\right] v_x^2 \nn \\
m_{3}^{2} &=& c_B f-\frac{f}{M^2} \left[ (c_1+c_2)\,\mu \,v_x-c_B(c_{1}+c_{2}+c_{X})v_x^2 \right]~.
\end{eqnarray}
Note that in the limit  where the sgoldstino decouples, due to the scaling $v_x\sim1/m_x^2$ in (\ref{EWxinEW}), $v_x\to 0$ which reproduces the usual MSSM expressions for these mass parameters.
The dimensionless coefficients in (\ref{Vtot}) are given  by
\begin{equation}
\label{la}
\lambda_{a}=\lambda_{a}^{(D)}+\lambda_{a}^{(F_X)}+\lambda_{a}^{(x)}~~~~~~~~~~~~a=1,\cdots,7
\end{equation}
which correspond, respectively,  to the contributions that arise as a consequence of integrating out the auxiliary $D$-components of the vector multiplets (i.e.\,the only contribution in the MSSM),
\begin{eqnarray}
\lambda_{1}^{(D)} & =&  \lambda_{2}^{(D)}= \frac{g_{2}^{2}+g_{1}^{2}}{4}~,~\lambda_{3}^{(D)}= \frac{g_{2}^{2}-g_{1}^{2}}{4}~,~\lambda_{4}^{(D)}= -\frac{g_{2}^{2}}{2}\nn \\
\lambda_{5}^{(D)}&=&\lambda_{6}^{(D)}=\lambda_{7}^{(D)}=0 
\end{eqnarray}
 the F-component and the goldstino multiplet,
\begin{eqnarray}
\lambda_{1}^{(F_X)}&=& 2c_{1}^{2}\frac{f^2}{M^4}  ~,~\lambda_{2}^{(F_X)}=2c_{2}^{2}\frac{f^2}{M^4}~,~\lambda_{3}^{(F_X)}=2c_{1}c_2\frac{f^2}{M^4} ~,~\lambda_{4}^{(F_X)}=c_B^2 \nn \\
\lambda_{5}^{(F_X)}&=&0~,~ \lambda_{6}^{(F_X)}=-c_B c_{1} \frac{f}{M^2}~,~ \lambda_{7}^{(F_X)}=-c_B c_{2} \frac{f}{M^2}~.
\end{eqnarray}
and the heavy sgoldstino scalar,
\begin{eqnarray}
\lambda_{1}^{(x)} &=& \lambda_{2}^{(x)}= \lambda_{3}^{(x)} =- 2\frac{c_B^2\,\mu^2 M^2 }{c_X f^2}~,~
 \lambda_{4}^{(x)}  = -\frac{\mu^2 (c_{1}+c_{2})^2}{c_X M^2}\nn  \\
\lambda_{5}^{(x)} & = & 0~,~\lambda_{6}^{(x)} =  \lambda_{7}^{(x)} =\frac{c_B \mu^2 (c_{1}+c_{2}) }{c_X f }  ~.
\end{eqnarray}
which arise when couplings of the form $xhh$ in (\ref{VFh2}) are connected with an sgoldstino propagator, in analogy with the discussion in \cite{Brignole:2003cm}. In contrast to the MSSM, where the tree level coefficients $ \lambda_{6}$ and $\lambda_{7}$ are zero, there are here contributions to these coefficients due to the dynamical treatment of the goldstino multiplet. Hence, this treatment gives rise to leading order effects in terms of new types of quartic Higgs couplings. 

The Higgs VEV can be related to the mass parameters and the dimensionless coefficients in the following way
\begin{equation}
\label{vsquare}
v^2=-\frac{m^2}{\lambda}
\end{equation}
where 
\begin{equation}
m^2=\widetilde{m}_{1}^{2}\cos^2\beta+\widetilde{m}_{2}^{2}\sin^2\beta-m_3^2\sin2\beta
\end{equation}
\begin{eqnarray}
\label{l}
\lambda&=&\frac{\lambda_1}{2}\cos^4\beta+\frac{\lambda_2}{2}\sin^4\beta+(\lambda_3 +\lambda_4 +\lambda_5 )\cos^2\beta\sin^2\beta\nn \\
&&+2\lambda_6 \cos^3\beta\sin\beta+2\lambda_7\cos\beta\sin^3\beta~.
\end{eqnarray}
In analogy with (\ref{la}) we can separate the different contributions to (\ref{l}),
\begin{equation}
\lambda=\lambda^{(D)}+\lambda^{(F_X)}+\lambda^{(x)}
\end{equation}
where
\begin{eqnarray}
\lambda^{(D)}&=&\frac{g_{2}^{2}+g_{1}^{2}}{8}\cos^2 2\beta \label{lD}\\
\lambda^{(F_X)} & = & \left( c_{1}\frac{f}{M^2}\cos^2 \beta+c_{2}\frac{f}{M^2}\sin^2 \beta-\frac{c_B}{2}\sin2\beta \right)^2   \nn \\
&=& \left( \frac{\mu^2}{f}-\frac{c_B}{2}\sin2\beta \right)^2 \label{lFx}\\
\lambda^{(x)}&=& -\frac{\mu^2 M^2}{c_X f^2}\left( c_{B}\cos^2 \beta+c_{B}\sin^2 \beta-\frac{(c_1+c_2)}{2M^2}\sin2\beta \right)^2 \nn \\
&=& -\frac{\mu^6 }{m_{x}^2 f^2} \sin^2 2\beta  \label{lx}
\end{eqnarray}
where in the second line of (\ref{lFx}) and (\ref{lx}) we have imposed the minimization conditions in (\ref{EWh1}) and (\ref{EWh2}). The contributions in (\ref{lD}),  (\ref{lFx}) and  (\ref{lx}) are the ones appearing in the expressions for the mass for the lightest Higgs particle in (\ref{mh}) (with a multiplicative factor $4v^2$). In the MSSM, the total tree level contribution to $\lambda$ in (\ref{l}) is given by (\ref{lD}) and it is the smallness of $\lambda^{(D)}$ that is the origin of the little hierarchy problem. The positive contribution in (\ref{lFx}) increases $\lambda$ while the negative contribution in (\ref{lx}) decreases it, as discussed in Section 3.

 \section{Corrections in the Light Sgoldstino Case}
 
We give here more details concerning the higher order terms in $C_X$, $\gamma$ and $\tilde{\gamma}$  for the EW symmetry breaking conditions and the neutral scalar masses given in Section 4.

\subsection{Large $\tan \beta $ }
 
 Referring to the notation used in the main text, at the first order in $C_X$ and the second in $\gamma=(\tan \beta)^{-1}$ and $v$, we have\footnote{Since $C_X$ and $\gamma$ are both considered to be small, we neglect  the order $C_X \gamma^2$.}
\begin{eqnarray}
&& c_{1,x} =  \frac{2 \mu ^2 \left(4 v^2 \mu ^4+2 f^2 \left( m_Z^2- \mu ^2\right)\right)
- C_B \left(6 f v^2
\mu ^4+2 f^3 \left(m_Z^2- \mu ^2\right)\right)}{16 f \mu ^6} \nonumber \\
&& c_{1,\gamma}= \frac{1}{32 f \mu ^4} \Big[4 C_B^2  \left(f^2 m_Z^2 +  \mu ^4\right)  \nonumber \\
&& \qquad \qquad+ C_B  \frac{\mu^2}{f} \left(-78 v^2
\mu ^4+f^2 \left(34 m_Z^2-20 \mu ^2\right)\right) \nonumber \\
&& \qquad \qquad+ 2 \frac{\mu^4}{f^2} \left(-6 v^2 \mu^4+2 f^2 \left(3 m_Z^2-2 \mu ^2\right)\right) \Big] \nonumber \\
&& c_{2,x} = 0  \nonumber \\
&& c_{2,\gamma}= \frac{2\mu ^2 \left(f^2 m_Z^2 + 4 v^2 \mu ^4\right)+2 C_B \left(12 f v^2 \mu ^4-2 f^3 \left(m_Z^2-2 \mu ^2\right)\right)}{4 f^3 \mu ^2}~.
\end{eqnarray}
The constraint on $C_X$, arising from demanding that these first order corrections are smaller than the leading order term, is of the same order as the one  considered in Section 4. One should also consider a similar constraint on $\gamma$, but the region for $\tan\beta$ considered in the analysis is well inside the allowed one. \\
Concerning the corrections to the masses introduced above, one obtains
\begin{eqnarray}
&& \delta_{\mathrm{Re}x, C_X} =\frac{C_B f v^2 (C_B f+6 \mu^2)}{(C_B f+2 \mu^2)^2}  \nonumber \\
&& \delta_{\mathrm{Re}x,\gamma} = -\frac{2 v^2  \mu^2 (4 C_B^3 f^3+2 C_B^2 f^2 \mu^2+9 C_B f \mu^4+2 \mu^6)}{f^2 (C_B f+2 \mu^2)^2} \nonumber \\
&& \delta_{H^0, C_X} = \frac{1}{8} \Big[\frac{f^2 (m_Z^2-\mu^2)(2 \mu^2-C_B f)}{ \mu^6}-\frac{2 v^2 (2 C_B^3 f^3+5 C_B^2 f^2 \mu^2+4 C_B f \mu^4-28 \mu^6)}{(C_B f \mu+2 \mu^3)^2}\Big] \nonumber \\
&& \delta_{H^0, \gamma} = \frac{1}{16} \Big[ -\frac{80~ C_B^4 f^2 v^2}{(C_B f+2 \mu^2)^2}+\frac{124 C_B^3 f v^2}{C_B f+2 \mu^2}-\frac{6 C_B^2 f^2 m_Z^2}{\mu^4}+\frac{C_B f (8 C_B f+41 m_Z^2)}{\mu^2} \nonumber \\
&& \qquad \qquad \qquad +\frac{\mu^2 (35 C_B v^2-4 f)}{f}-2 C_B (5 C_B v^2+13 f)-\frac{34 v^2 \mu^4}{f^2}+ 94 ~m_Z^2 \Big]\nonumber \\
&& \Delta_A = -\frac{2 (C_B f v^2 \mu^2+2 f^2 m_Z^2-2 v^2 \mu^4)}{f^2}~.
\end{eqnarray}

\subsection{Small $\tan \beta $ }

Again, using the notation used in the main text, the corrections to the EW symmetry breaking conditions (\ref{EWsmT1}), with $\tilde{\gamma} = \frac{1}{2} (\tan \beta -1)$, are given by
\begin{eqnarray}
&& \tilde{c}_{1,x} = \tilde{c}_{2,x} =  \frac{c_B v^2 (c_B f-\mu^2)}{(c_B f+\mu^2)^2}  \nonumber \\
&& \tilde{c}_{1,\tilde{\gamma}} =  \frac{c_B^3 f}{2 (c_B f+\mu^2)^5} \Big(c_B^2 f^3 (-2 c_B^2 v^2+4 c_B f+2 m_Z^2 )\nonumber \\
&& \qquad \qquad \qquad \qquad +2 c_B f^2 \mu^2 ( -c_B^2 v^2 +2 m_Z^2)+6 c_B f)\nonumber \\
&& \qquad \qquad \qquad \qquad+2 \mu^6 (c_B v^2+2 f)+2 c_B f \mu^4 (c_B v^2+6 f)\Big) \nonumber \\
&& \tilde{c}_{2,\tilde{\gamma}} = - \tilde{c}_{1,\tilde{\gamma}} 
\end{eqnarray}
whereas for the masses one obtains
\begin{eqnarray}
&& \tilde{\delta}_{\mathrm{Re}x, C_X} = v^2 \Big(1-\frac{6 \mu^2}{c_B f} \Big)   \nonumber \\
&& \tilde{\delta}_{\mathrm{Im}x, C_X} =  v^2 \Big(1 + \frac{2 \mu^2}{c_B f} \Big)  \nonumber \\
&& \tilde{\delta}_{H^0, C_X}= -\frac{2 (7 c_B^3 f^2 v^2+2 c_B^2 (2 f^3+5 f v^2 \mu^2)+c_B (3 v^2 \mu^4-f^2 (m_Z^2-8 \mu^2))+4 f \mu^4)}{c_B^2 f (c_B f+\mu^2)}  \nonumber \\
&& \tilde{\delta}_{A^0, C_X}= -\frac{4 (7 c_B^2 f v^2+2 c_B f^2+11 c_B v^2 \mu^2+2 f \mu^2)}{c_B^2 f}   \nonumber \\
&& \tilde{\delta}_{h^0, C_X}= -\frac{2 v^2 (3 c_B f+5 \mu^2)}{c_B f}~.
\end{eqnarray}

\end{document}